\documentclass[conference]{ieeeconf}

\usepackage{amsmath}
\usepackage[dvipsnames]{xcolor}
\usepackage[hidelinks]{hyperref}
\usepackage{graphicx}
\usepackage{amsthm}
\usepackage{algorithm}
\usepackage{algpseudocode}
\usepackage{tikz}

\usepackage[normalem]{ulem}

\newcommand\redout{\bgroup\markoverwith{\textcolor{red}{\rule[.5ex]{2pt}{0.4pt}}}\ULon}

\newcommand{\Vast}{\bBigg@{5}}

\newtheorem{lemma}{Lemma}
\newtheorem{theorem}{Theorem}
\newtheorem{proposition}{Proposition}
\newtheorem{assumption}{Assumption}
\newtheorem{example}{Example}

\title{Throughput-Optimal Scheduling via Rate Learning}

\author{Panagiotis Promponas, V\'ictor Valls$^*$, Konstantinos Nikolakakis,\\ Dionysis Kalogerias, and Leandros Tassiulas\vspace{2bp}\\
Yale University, $^*$IBM Research Europe -- Dublin}

\begin{document}

\maketitle

\begin{abstract}
We study the problem of designing scheduling policies for communication networks. This problem is often addressed with max-weight-type approaches since they are throughput-optimal. However, max-weight policies make scheduling decisions based on the network congestion, which can be sometimes unnecessarily restrictive. In this paper, we present a ``schedule as you learn'' (SYL) approach, where we learn an average rate, and then select schedules that generate such a rate in expectation. This approach is interesting because scheduling decisions do not depend on the size of the queue backlogs, and so it provides increased flexibility to select schedules based on other criteria or rules, such as serving high-priority queues. We illustrate the results with numerical experiments for a cross-bar switch and show that, compared to max-weight, SYL can achieve lower latency to certain flows without compromising throughput optimality. 
\end{abstract}

\section{Introduction}
\label{sec:introduction}

Scheduling policies in networking have historically focused on maximizing throughput, as this was the main bottleneck in communication networks. This is, however, not the case anymore with modern communication networks (e.g., 5G), and cloud applications (e.g., collaborative LaTeX editors and online video games) often require performance metrics besides throughput to achieve a good quality of experience. For example, a latency above 250 ms is considered not suitable for online gaming \cite{WCH+21}.

\begin{figure}
\centering
\includegraphics[width=0.75\columnwidth]{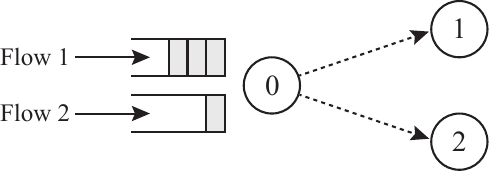}
\vspace{-0.15cm}
\caption{Server with two queues. The server can only serve one packet from the queues at a time. Flow 1 and 2 are directed to nodes 1 and 2, respectively.}
\label{fig:toy_network}%
\vspace{-6bp}
\end{figure}

\begin{figure}
\includegraphics[width=\columnwidth,height=0.2\textheight,keepaspectratio]{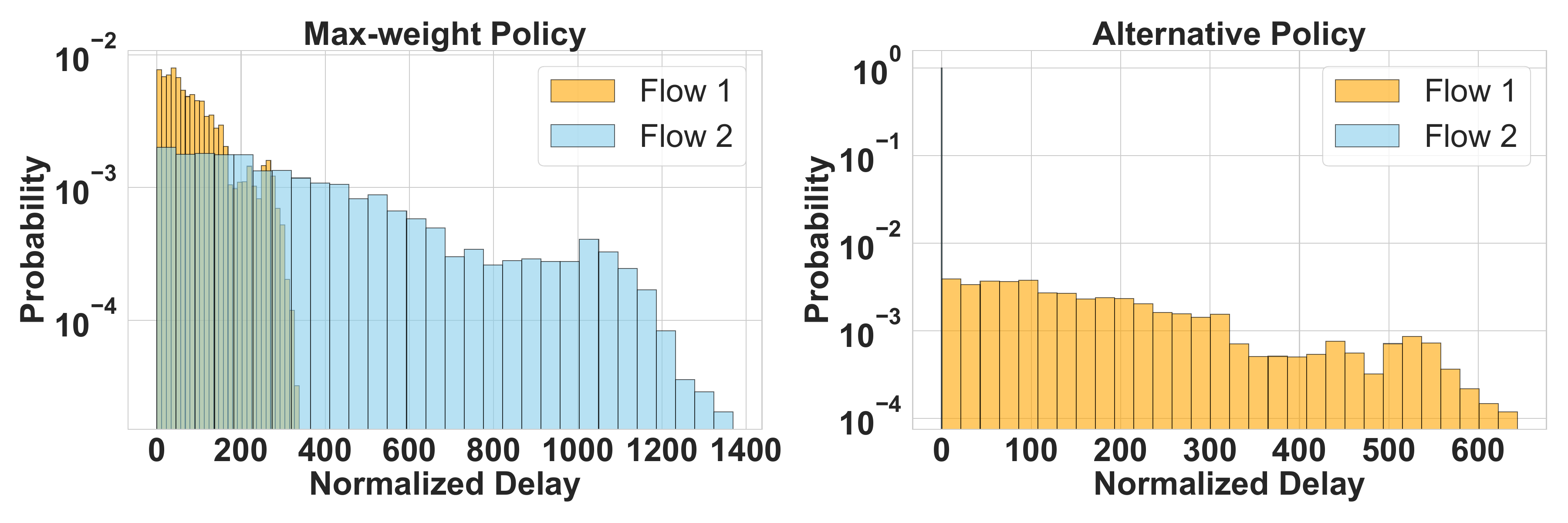}
\vspace{-0.7cm}
  \caption{Probability distribution of the packets waiting times for the network in Fig.~\ref{fig:toy_network}. The waiting times are normalized to the frequency in which the server selects schedules. The packet arrivals of flow 1 and flow 2 are Bernoulli with mean $(1-\epsilon)0.8$ and $(1-\epsilon) 0.2$ respectively, where $\epsilon = 5\cdot 10^{-5}$. }
  \label{fig:toy_histograms}
\vspace{-12bp}
\end{figure}

Let us illustrate, with a toy example, the problem of a scheduling policy that focuses only on maximizing throughput. Consider the network illustrated in Fig.~\ref{fig:toy_network}; a server that receives packets from two different flows, and stores them in separate queues. Flow 1 represents a flow that is insensitive to latency (e.g., FTP traffic), and flow 2 is a flow with low-latency requirements (e.g., an online video game). In this example, the server can transmit at most one packet at a time, so it has to decide which of the queues to serve.  In this case, the max-weight policy \cite{TE92} will transmit a packet from the queue with the largest backlog. Such a strategy is throughput-optimal in the sense that all the packets that get into the queues will eventually get out in a finite amount of time.\footnote{Or more formally, the waiting times of the packets in the queues does not grow unbounded.} However, the max-weight policy does not provide low latency to flow 2 (see Fig.~\ref{fig:toy_histograms}). Now, consider a policy that transmits a packet from queue 2 if this is not empty, and otherwise it serves a packet from queue 1. This \emph{alternative policy} is also throughput optimal, but this provides, in addition, low latency to the second flow (see Fig.~\ref{fig:toy_histograms}). Hence, it is preferable over max-weight for this scenario.

Designing throughput optimal policies is highly non-trivial, especially when the network has arbitrary topology, the traffic arrival statistics are not known, and the network connectivity varies over time. The max-weight policy \cite{TE92} and variants \cite{GNT06, Nee10} have become the \emph{de facto} approaches for maximizing throughput in networks. Their fundamentals are well studied by the networking and control communities, and besides throughput-optimality, they have the appealing property that scheduling decisions can be computed in polynomial time by finding a maximum-weighted matching in a graph \cite{shoemaker2016edmonds}. 
In addition, well-established extensions of max-weight, such as the Lyapunov Drift-Plus-Penalty (LDPP) algorithm \cite{Nee10}, can minimize a convex utility function of the average throughput. For example, LDPP can serve all the traffic in a wireless network while minimizing the energy used \cite{neely2015energy}.
However, performance metrics, such as latency, are not a function of the average flow, and selecting schedules to be maximum-weighted matchings in a graph may not always be the best choice, as shown in the example in Fig.~\ref{fig:toy_network}.

In this paper, we study the problem of designing throughput optimal scheduling policies for networks with arbitrary topology when the statistics of the arrival process are unknown. Unlike max-weight approaches, which seek to stabilize a system of queues directly, we see the problem as learning an ``average'' rate vector that guarantees queue stability, and then ``stack'' a (randomized) scheduling policy on top that guarantees such an average rate in expectation. That is, \textit{we select schedules as we learn an average rate}. Such an approach is appealing because it decouples the choice of schedule from the size of the queue backlogs at a given time, thus providing more flexibility as to \textit{how} to select schedules. Max-weight is \emph{a} throughput-optimal policy, but not the only one. Among all throughput-optimal policies, it is interesting to select/design one that boosts other performance metrics important for the problem of interest, e.g., low latency or reduced memory usage. To this end, the main contributions of the paper are: 
\begin{itemize}
\item A new approach to designing throughput optimal policies
(Sec.~\ref{sec:formulation}). The strategy consists of learning an average rate that guarantees queue stability/maximum throughput, and then selecting schedules that ensure such an average rate in expectation. 
\item A new algorithm (Algorithm \ref{alg:new_algorithm}) that is throughput optimal. The algorithm uses Nesterov's dual-averaging \cite{Nes09} for learning the ``average'' rate, and a randomized policy for selecting the schedules. In our approach, scheduling decisions do not depend on the size of the queue backlogs. 

\item  Two numerical experiments for a cross-bar switch scheduling problem that illustrate the performance of our approach. In addition, we present a variant of SYL to illustrate how we can ``bias'' the selection of schedules to reduce the latency of a flow while retaining throughput optimally. 




\end{itemize}

Our approach is novel and provides an alternative perspective on how to tackle queue scheduling problems. Furthermore, we believe it makes the design of \emph{new} throughput-optimal policies conceptually simpler. The learning process can be offloaded to a subroutine, which allows us to focus on designing a scheduling policy for a particular system. Such an approach can be helpful for systems with complex operational constraints, such as quantum networks with qubit memory constrains and decoherence \cite{10186423}. 


The rest of the paper is organized as follows. Sec.~\ref{sec:related_work} gives an overview of previous work and position our contributions. Sec.~\ref{sec:preliminaries} introduces the preliminaries: the notation, a sufficient condition for queue stability, and the dual averaging algorithm. In Sec.~\ref{sec:formulation}, we formulate the problem and discuss how we can tackle it depending on whether the statistics of the packet arrivals are known, or not. Sec.~\ref{sec:main_results} presents the main technical results (Algorithm \ref{alg:new_algorithm} and Theorem \ref{th:main_theorem}) and discusses the limitations of our approach. Finally, Sec.~\ref{sec:numerical_experiments} presents the numerical experiments for a cross-bar switch, and Sec.~\ref{sec:conclusions} concludes. All the proofs are in the Appendix. 


\section{Background and Related Work}
\label{sec:related_work}

A significant amount of research on throughput-optimal policies started as result of the paper in \cite{TE92}. In brief, \cite{TE92} characterized the set of rates for which throughput-optimal policies exist and showed how to design a throughput-optimal policy using a discrete-time Lyapunov control approach. The throughput-optimal policy presented in \cite{TE92} is known as max-weight, as this selects schedules that correspond to maximum-weighted matchings in the connectivity graph---the weights of edges are the sizes of the queue backlogs. This line of work has inspired several extensions of the original algorithm with applications to many networking and resource allocation problems. Some notable extension include utility functions \cite{Nee10}, constrained schedules 
\cite{celik2013scheduling}, heavy-tailed traffic \cite{7161409}, and online learning \cite{neely2017online}.

The appealing features of max-weight algorithms\footnote{Throughput optimality; it does not require to know the statistics of the arrival process; and schedules can be computed in polynomial time.} motivated researchers to develop convex optimization algorithms for network resource allocation problems with similar features. In short, the approaches often consists of connecting Lagrange multipliers with scaled queues' occupancy in a specific convex optimization method, and then show, via the Lagrange multipliers, that the queues are stable. See, for example,  
\cite[Sec.~4.7]{huang2011deterministic},  \cite[Sec.~V]{8449113}, \cite[Sec.~IV-B]{7874210}. 

The approach in this paper is conceptually different from max-weight and convex optimization approaches. We show that satisfying an \emph{average rate} vector condition is sufficient for obtaining queue stability and then aim to find/learn such a rate. We use dual-averaging (i.e., a convex optimization algorithm) to learn the rate vector, but other techniques are possible. Notably, our learning algorithm does not connect dual variables with queues, which is the distinctive characteristic of previous convex optimization works \cite{huang2011deterministic,8449113,7874210}.  Furthermore, we use \emph{diminishing} step sizes to learn the rate vector (i.e., $\alpha \to 0$), which is in contrast to the \emph{constant} step size employed in previous approaches  to ensure queue stability (see, for example, \cite[Sec.~4.10]{GNT06} and \cite[Sec.~V]{8449113}).

\section{Preliminaries}
\label{sec:preliminaries}

This section introduces the notation, a sufficient condition for strong stability in queuing systems, and the dual-averaging algorithm that we will use in Sec.~\ref{sec:main_results} to learn a rate vector that ensures queue stability. 

\subsection{Notation}

We use $\mathbf R^n$ to denote the set of $n$-dimensional vectors, and $\mathbf R^n_+$ the set of vectors with non-negative entries. All vectors are in column form, and we use $\mathbf 1$ to denote the all ones column vector---its dimension will be clear from the context. For two vectors $x$ and $y$, we use $\langle x, y \rangle$ to denote their inner product. We use $x \preceq (\succeq) y$ to indicate that $x$ is component-wise smaller (greater) than $y$. The function $[x]^+ : \mathbf R^n \to \mathbf R^n$ returns a vector where its $j$-th component is equal to the $j$-th component of $x$ if this is greater than or equal to zero, or zero otherwise. Finally, we use $\| \cdot \|$ to denote the $\ell_2$-norm, and $\nabla f(x)$ to denote a (sub)gradient of a function $f$ at $x$.

\subsection{Strong stability of a queuing system}
\label{sec:strong_stability}
Consider a system of $n$ queues that operates in slotted time. In each time slot $k=1,2,\dots$, the queues evolve according to the following recursion: 
\begin{align}
Q_{k+1} = \left[ Q_k + Z_k \right]^+ && k=1,2,\dots 
\label{eq:queue_update_preliminaries}
\end{align}
where $Q_1 \in \mathbf R^n_+$, and $Z_k \in \mathbf R^n$ indicates the net increment of items in each of the queues.  

A queuing system is strongly stable when the expected backlog of all queues is finite \cite{Nee10}, i.e., 
\begin{align}
\lim_{k\to \infty} \frac{1}{k} \sum_{i=1}^k \mathbf E[ \| Q_i \| ] \prec \infty.
\label{eq:strong_stability_ssq}
\end{align}
Strong stability is an important property to ensure throughput-optimality. Informally, by ensuring that the queues do not overflow, we guarantee that all the packets that get into the queues will eventually get out. 
A queueing system will be strongly stable depending on the properties of the sequence $\{Z_i \}_{i=1}^k$. The following proposition establishes a simple condition to have strongly stable queues. 

\begin{proposition}
Consider the update in Eq.~\eqref{eq:queue_update_preliminaries} with $Q_1 \in \mathbf R^n_+$. Suppose that 
\begin{align*}
    \mathbf E[Z_k] \preceq -\eta \mathbf 1 && \text{for all $k=1,2,\dots$}
\end{align*}
with $\eta > 0$ and $\| Z_k \| \le \sigma$ for some constant $\sigma < \infty$. Also, suppose that $Z_k$ is independent of $Q_k$. Then, the queuing system is strongly stable.
\label{prop:queue_stability}
\end{proposition}

The result above holds also when the condition $\mathbf E[Z_k] \preceq -\eta \mathbf 1$ is only satisfied for $k \ge \tau$, where $\tau \in \{1,2,\dots\}$ and $Q_\tau \in \mathbf R^n_+$. We will use this fact in Lemma \ref{th:tau_lemma} in Sec.~\ref{sec:main_results}.

\subsection{Dual averaging}
\label{sec:dual_averaging}

We present Nesterov's dual averaging  algorithm \cite{Nes09} in generic convex optimization form as this will allow us to streamline the proofs and exposition in Sec.~\ref{sec:main_results}.

Consider the convex optimization problem:
\begin{align}
\begin{tabular}{ll}
$\underset{x \in C}{\text{minimize}}$ & $f(x)$
\end{tabular}
\label{eq:primal_problem}
\end{align}
where $f : \mathbf R^n \to \mathbf R$ and $C \subseteq \mathbf R^n$ are convex. 
%
%
The objective function does not need to be differentiable. To minimize $f$, we generate a sequence $\{x_i \in C \}_{i=1}^k$ to query a stochastic first-order oracle. On querying the oracle at point $x_k$, this returns vectors of the form
\begin{align*}
\widetilde \nabla f(x_k) = \nabla f(x_k) + \xi_k,
\end{align*}
where $\xi_k$ is a bounded random variable with $\mathbf E[\xi_k] = 0$ for all $k$. With such an oracle, we can design an algorithm that solves \eqref{eq:primal_problem}. The following lemma is a special case of \cite[Theorem 1]{Nes09}, and we will use it to prove Lemma \ref{th:dual_dual_averaging} in Sec.~\ref{sec:main_results}.

\begin{algorithm}[t!]
\caption{Dual averaging algorithm}\label{alg:dual_averaging}
\begin{algorithmic}
\State \textbf{Input:} $f$ and $C$ convex, $\phi$ is $\sigma$-strongly convex
\State \textbf{Set:} $s_0 = 0$, $k = 1$
\While{termination condition is not met}
\State $\alpha_k > 0$
\State $x_{k} = \arg \max_{u \in C} \{ \langle s_{k-1}, u  \rangle - \phi(u) \}$
\State $s_{k} \leftarrow s_{k-1} - \alpha_k \widetilde \nabla f(x_k)$
\State $k \leftarrow k + 1$
\EndWhile
\end{algorithmic}
\end{algorithm}

\begin{lemma} Consider the optimization in \eqref{eq:primal_problem}. 
Let $\phi$ be a $\sigma$-strongly convex  function such that $\phi(u) \ge 0$ for all $u \in C$. Suppose that $\mathbf E[\widetilde \nabla f(x_k)] = \nabla f(x_k)$. For any $w \in C$, Algorithm \ref{alg:dual_averaging} ensures that

\begin{align}
&  \mathbf E \left[ \sum_{i=1}^k \alpha_i f(x_i) - k f(w) \right] \notag \\  
 & \qquad \le \mathbf E \left[ \sum_{i=1}^k \alpha_i \langle  \nabla f(x_i), x_i - w \rangle \right]  \label{eq:dual_averaging_lemma}
\notag \\
& \qquad \le  \phi(w)   + \frac{1}{2\sigma} \sum_{i=1}^k \alpha_i^2 \| \widetilde \nabla f(x_i) \|^2 .
\end{align}
\label{th:dual_averaging}
\end{lemma}
An example of function $\phi(u)$ is $\frac{1}{2} \| u - x_0 \|^2$, i.e., a function that measures the distance between $u$ and a vector  $x_0 \in C$. 
The bound in Eq.~\eqref{eq:dual_averaging_lemma} is closely connected to the convergence of subgradient methods \cite{BM08}. If we set $w = x^\star : = \arg \min_{u \in C} f(u)$ and assume that $\| \widetilde \nabla f(x_k)\|$ are uniformly upper bounded for all $k$, one can show that $f(x_k)$ converges, in probability, to $f(x^\star)$ (see \cite{BM08} for the details).


\section{Problem Formulation}
\label{sec:formulation}

Consider a system of queues that evolve as follows:
\begin{align*}
Q_{k+1} = \left[ Q_k + A_k  - S_k \right]^+ && k=1,2,\dots 
\end{align*}
where $Q_k \in \mathbf R_+^n$,  $\{ A_i \in \mathcal A \}_{i=1}^k$ and $\{ S_i \in \mathcal S\}_{i=1}^k$ are the packet arrivals and service in a time slot $t$. Without loss of generality, we assume that $\mathcal A : = \{a_1, \dots, a_{|\mathcal A|} \}$ and $\mathcal S:= \{ s_1, \dots, s_{|\mathcal S|} \}$ are discrete sets in $\mathbf R^n$, e.g., $\mathcal A, 
\mathcal S \subseteq \{0,1\}^n$. Also, we will refer to the elements in $\mathcal S$ as schedules. 

For a given arrival process $\{A_i \in \mathcal A\}_{i=1}^k$, our goal is to design a policy that generates a sequence $\{S_i \in \mathcal S\}_{i=1}^k$ that stabilizes the system of queues (see Sec.~\ref{sec:strong_stability}). Such a problem is feasible when
\begin{align*}
\lambda : = \lim_{k \to \infty} \frac{1}{k} \sum_{i=1}^k A_i, && A_i \in \mathcal A
\end{align*}
is in the set
\begin{align*}
\Lambda := \{ \lambda \in \mathbf R^n_+: \lambda \preceq \mu \ \text{for some } \mu \in \mathrm{conv}(\mathcal S) \},
\end{align*}
which is known in the networking literature as the \emph{network capacity region} \cite{TE92}. In short, the network capacity region contains all the vectors $\lambda$ such that $\lambda \preceq \mu$ for a vector $\mu$ in the convex hull of $\mathcal S$. Also, note that, by construction, we can write any vector $\mu \in \mathrm{conv}(\mathcal S)$ as the running average of service vectors in $\mathcal S$, i.e., $\mu  = \lim_{k \to \infty} \frac{1}{k} \sum_{i=1}^k S_i$, with $S_i \in \mathcal S$.

In the rest of the paper, we will make the following assumption, which ensures that the problem is \emph{strictly} feasible (see, for example, \cite[pp.~104]{GNT06}). 

\begin{assumption}
\label{as:slater}
$\lambda$ is in the interior of the network capacity region. That is, for a given vector $\lambda$, there exists a vector $\mu \in \mathrm{conv}(\mathcal S)$ and a $\eta > 0$ such that $\lambda + \eta \mathbf 1 \preceq \mu$.
\end{assumption}

%


The problem of designing a service policy that stabilizes the queuing system can be divided into two cases, depending on whether $\lambda$ is known, or not. We are interested in solving the problem when $\lambda$ is not known (Sec.~\ref{sec:lambda_not_known}), but present the case where $\lambda$ is known since both cases are closely connected in our work.

\subsection{$\lambda$ is known}
\label{sec:lambda_known}
In this case, we can design a service policy in two steps:
\begin{itemize}
    \item [(i)] Find a vector $\mu \in \mathrm{conv}(\mathcal S)$ such that $\lambda + \eta \mathbf 1 \preceq \mu$ for some $\eta > 0$.
    \item [(ii)] Decompose $\mu$ as the convex combination of elements in $\mathcal S$, i.e., write $\mu = \sum_{j=1}^{|\mathcal S|} \theta_j s_j$ for some $\theta_j \ge 0$ with $\sum_{j=1}^{|\mathcal S|} \theta_j = 1$.  Such a decomposition is always possible because $\mu$ is, by construction, in the convex hull of $\mathcal S$. 
\end{itemize} 

With the decomposition of $\mu$, we can select an element in $\mathcal S$ with probability equal to its weight, i.e., 
\begin{align}
\mathrm{Prob}(S_k = s_j) = \theta_j.
\end{align}
Such a policy is often known as the \emph{randomized policy}; it will generate a sequence $\{ S_i \in \mathcal S\}_{i=1}^k$ where $\mathbf E[S_k] = \mu $ for all $k \in \{1,2,\dots\}$. Thus, we have that $\mathbf E[A_k] - \mathbf E[S_k] \preceq -\eta \mathbf 1$ and so the queues will be strongly stable by Proposition \ref{prop:queue_stability}. 

 Finding a vector $\mu$ that ensures $\lambda + \eta \mathbf 1 \preceq \mu$ is easy if we can check whether a vector is in the convex hull of $\mathcal S$. For instance, we could solve the optimization $\max_{\eta \ge 0} \eta$ s.t.\ $\lambda + \eta \mathbf 1 \in \mathrm{conv}(\mathcal S)$ with the bisection method. The complexity of decomposing a vector $\mu$ as a convex combination of vectors in $\mathcal S$ depends on the problem. For example, if the number of elements in $\mathcal S$ is small, we can obtain such a decomposition by solving the convex optimization problem: $\min_{\theta_j \ge 0} \| \sum_{j=1}^{|\mathcal S|} \theta_j s_j - \mu \|^2$ s.t.\ $\sum_{j=1}^{|\mathcal S|} \theta_j = 1$.  However, that is not always possible. For example, in the cross-bar switches, the number of actions (i.e., $|\mathcal S|$) increases factorially with the number of input/output ports \cite{bojja2016costly}.\footnote{e.g., a switch with 16 ports has over 20 trillion schedules/actions.} 
 
To conclude this section, we present an example of how to decompose a $3\times 3$ traffic matrix of a cross-bar switch with the approach described above. 

\begin{example} Consider a $3 \times 3$ cross-bar switch with arrival matrix (i.e., a 9-dimensional vector):
\begin{align}
\lambda = 
\begin{bmatrix}
 0.6 &  0.3 &  0.0 \\
 0.1 &  0.0 & 0.8 \\
 0.2 & 0.6 & 0.1 
\end{bmatrix}.
\label{eq:lambda}
\end{align}
Set $\mathcal S$ corresponds to the permutation matrices plus the null action (i.e., the all-zeros matrix). Matrix $\lambda$ is in the interior of the capacity region since the sum of the rows and columns is equal to 0.9. A vector $\mu$ that satisfies $\lambda + \eta \mathbf 1 \preceq \mu$ is 
\begin{align*}
\mu = 
\begin{bmatrix}
 0.633 &  0.333 &  0.033 \\
 0.133 &  0.033 & 0.833 \\
 0.233 & 0.633 & 0.133 
\end{bmatrix}.
\end{align*}
That is, we have added $0.033$ to each entry of $\lambda$. We can decompose $\mu$ as follows:
\begin{align*}
 \frac{19}{30} \left[\begin{smallmatrix}
 1 & 0 & 0 \\
 0 & 0 & 1 \\
 0 & 1 & 0 
\end{smallmatrix}\right]
+ \frac{6}{30}\left[\begin{smallmatrix}
 0 & 1 & 0 \\
 0 & 0 & 1 \\
 1 & 0 & 0 
\end{smallmatrix} \right] 
+ \frac{4}{30} \left[\begin{smallmatrix}
 0 & 1 & 0 \\
 1 & 0 & 0 \\
 0 & 0 & 1 
\end{smallmatrix} \right]
 + \frac{1}{30} \left[\begin{smallmatrix}
 0 & 0 & 1 \\
 0 & 1 & 0 \\
 1 & 0 & 0 
\end{smallmatrix}\right].
\end{align*}
A randomized policy will select each permutation matrix, respectively, with probability $\frac{19}{30}$, $\frac{6}{30}$, $\frac{4}{30}$, and $\frac{1}{30}$. 
\label{example_numerical}
\end{example}

\subsection{$\lambda$ is not known}
\label{sec:lambda_not_known}

A straightforward approach to tackle this case is to observe the arrivals process $\{A_i \in \mathcal A\}_{i=1}^k$ for long enough to ensure that $\frac{1}{k} \sum_{i=1}^k A_i$ is sufficiently close to $\lambda$ to estimate a vector $\mu$ such that $\lambda + \eta \mathbf 1 \preceq \mu$ for some $\eta > 0$. With that, we can decompose $\mu$ as indicted in the previous section. However, such an approach is typically not desirable because it requires to remain idle (i.e., not serve packets) for a long period of time.

An alternative approach is the following. Suppose that we have access to a sequence $\{ \bar \mu_i \in \mathrm{conv}(\mathcal S)\}_{i=1}^k$ such that $\lambda + \eta \mathbf 1 \preceq \mathbf E[ \bar \mu_k ] $ for all $k \ge \tau$ for some constant $\tau \in \{1,2,\dots \}$. Then, we can use $\bar \mu_k$ to construct a ``randomized'' policy as explained in Sec.~\ref{sec:lambda_known}. In particular, at time $k$, we will write $\bar \mu_k$ as the convex combination of elements in $\mathcal S$, and select one of the elements with probability proportional to the weights. The difference with the approach mentioned in the paragraph above is that we do not wait to learn a $\mu$ that satisfies $\lambda + \eta \mathbf 1 \preceq \mu$ to start the scheduling process. Instead, we schedule as we learn. In the next section, we present an algorithm that learns $\bar \mu_k$ with Nesterov's dual averaging algorithm but other optimization techniques could be used as long as the necessary conditions are met.

\section{Main Results}
\label{sec:main_results}

\begin{algorithm}[t!]
\caption{Dual averaging algorithm with stochastic (sub)gradients for the Lagrange dual problem \eqref{eq:dual_problem}}\label{alg:dual_averaging_for_dual_problem}
\begin{algorithmic}
\State \textbf{Input:} $f$ and $C$ convex, $\phi (u) = \frac{1}{2} \| u\|^2$ 
\State \textbf{Set:} $s_1 = 0$, $x_1 \in C$, $k = 1$
\While{termination condition is not met}
\State $\alpha_k \leftarrow 1/ \sqrt k $
\State $y_{k} = \arg \max_{v \succeq 0} \{ \langle s_{k-1}, v \rangle - \phi(v) \}$

\State $\mu_k = \arg \min_{u \in C} \{ f(u) + \langle y_k, \lambda - u \rangle \}$
\State $s_{k+1} \leftarrow s_{k} +  \alpha_k (A_k - \mu_k)$

\State $k \leftarrow k + 1$
\EndWhile
\end{algorithmic}
\label{al:dual_dual_problem}
\end{algorithm}

In this section, we present an approach for constructing a sequence $\{\bar \mu_i \in C\}_{i=1}^k$ that satisfies $\lambda + \eta \mathbf 1 \preceq \mathbf E[ \bar \mu_k ] $ for all $k \ge \tau$ for some constant $\tau \in \{1,2,\dots \}$.

\begin{figure}
\centering
\includegraphics[width=0.9\columnwidth]{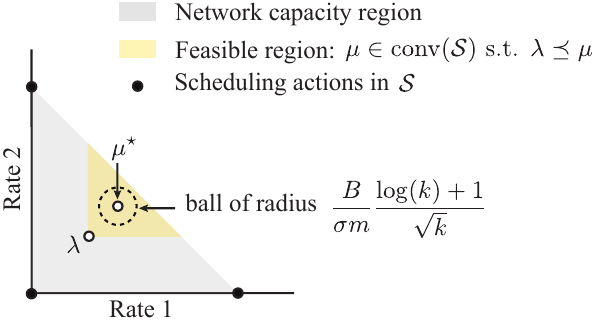}
\caption{Schematic illustration of the convergence result in Lemma \ref{th:dual_dual_averaging}. Vector $\bar \mu_k$ is within a ball of radius $B(\log(k) + 1)/(\sigma m \sqrt{k})$ centered at $\mu^\star$. By construction, $\mu^\star$ is strictly larger than $\lambda$.}
\label{fig:capacity_region}
\end{figure}

To start, consider the constrained optimization problem:
\begin{align}
\begin{tabular}{ll}
$\underset{\mu \in C}{\text{minimize}}$ & $f(\mu)$\\
subject to & $\lambda \preceq \mu$
\end{tabular}
\label{eq:constrained_problem}
\end{align}
where $\lambda$ is the average of the arrivals, $\mu$ the average service, and $C:=\mathrm{conv}(\mathcal S)$. Without loss of generality, we assume that $0 \in C$, i.e., there is the option to not serve any packets from the queues.\footnote{We will use the fact that $0 \in C$ in the proof of Lemma \ref{th:dual_dual_averaging}.} The objective function just expresses a preference on the average service rate $\mu$ that satisfy the constraints. In the schematic example in Fig.~\ref{fig:capacity_region}, the objective $f$ controls where $\mu^\star:= \arg \min_{\mu\in C, \lambda \preceq \mu} f(\mu)$ will be in the yellow region (i.e., the set of vectors where $\mu \succeq \lambda$). 

To tackle \eqref{eq:constrained_problem}, 
we formulate the Lagrange dual problem, which allow us to relax the knowledge of $\lambda$ by using stochastic (sub)gradients. This will be clear shortly. For now, consider the Lagrange dual problem of \eqref{eq:constrained_problem}:
\begin{align}
\begin{tabular}{ll}
$\underset{y \succeq 0}{\text{maximize}}$ & $h(y)$
\end{tabular}
\label{eq:dual_problem}
\end{align}
where 
\begin{align*}
    h(y):= \min_{\mu \in C} \{ f(\mu) + \langle y, \lambda - \mu\rangle \}.
\end{align*}
The Lagrange dual function $h$ is concave \cite[Chapter 5]{BV04}. Thus, we can use Algorithm \ref{alg:dual_averaging} since $-h$ is convex.

Next, consider that we minimize $-h$ by querying a stochastic first-order oracle at $y_k$ (see Sec.~\ref{sec:dual_averaging}). The oracle returns vectors of the form 
\begin{align*}
- \widetilde \nabla h(y_k) 
& = -\nabla h(y_k) + \xi_k  \\
& = \mu_k - \lambda +  \xi_k && (-\nabla h(y_k) = \mu_k - \lambda) \\
& = \mu_k - A_k  && (\xi_k := \lambda - A_k)
\end{align*}
where 
\begin{align}
    \mu_k & \in \arg \min_{u \in C} \{f(u) + \langle y_k, \lambda - u \rangle \}.  
    \label{eq:compute_mu}
\end{align}  
There are two important points to note. First, if $\mathbf E[A_k] = \lambda$, we have that $\mathbf E[\xi_k] = 0$ for all $k$, and so we satisfy the condition for dual averaging with stochastic gradients (see Sec.~\ref{sec:dual_averaging}). The second point is that we do not need to know $\lambda$ to compute $\mu_k$; see Eq.~\eqref{eq:compute_mu}. Thus, if $\mathbf E[A_k] = \lambda$, we can construct such an oracle by solving a convex optimization and ``replacing'' $\lambda$ with $A_k$. 
We have the following result. 

\begin{lemma} Suppose $f$ is $m$-strongly convex and that Slater's condition holds, i.e., there exists a $\eta >0 $ such that $\lambda + \eta \mathbf 1 \preceq \mu$ for some $\mu \in C$. Algorithm \ref{alg:dual_averaging_for_dual_problem} (i.e., Algorithm \ref{alg:dual_averaging} applied to problem \eqref{eq:dual_problem}) with $\phi(u) = \frac{1}{2} \| u\|^2$ ensures:
\begin{align*}
 \mathbf E \left[ \left\|  \bar \mu_k  - \mu^\star \right\|^2 \right]   
 \le \frac{B}{\sigma m} \frac{\log(k) + 1}{\sqrt{k}}
\end{align*}
where $\mu^\star:= \arg \min_{\mu\in C, \lambda \preceq \mu} f(\mu)$,
\begin{align}
    \bar \mu_k : = \frac{1}{\sum_{i=1}^k \alpha_i} \sum_{i=1}^k \alpha_i \mu_i,
    \label{eq:barmu}
\end{align} 
and $B:= \frac{1}{2} \max_{\mu \in C, a \in \mathcal A} \| \mu -a  \|^2 $. 
\label{th:dual_dual_averaging}
\end{lemma}
The result ensures that $ \mathbf E [ \left\|  \bar \mu_k  - \mu^\star \right\|^2 ] \to 0$ as $k\to \infty$ at a rate of $\tilde O(1/\sqrt k)$. Thus, for large enough $k$,  $\bar \mu_k$ will be within a ball around $\mu^\star$ with high probability. Fig.~\ref{fig:capacity_region} shows, schematically, the intuition behind this idea. The ball where $\bar \mu_k$ lives shrinks with $k$, and therefore, for $k$ sufficiently large, it will eventually happen that $\lambda \prec \mathbf E[\bar \mu_k]$ if $\lambda \prec \mu^\star$. The following lemma establishes a time $\tau \in \{1,2,\dots\}$ for which  $\mathbf E[\bar \mu_k]$ will be strictly larger than $\lambda$ for all $k\ge \tau$.

\begin{lemma}Suppose that $\lambda + \eta \mathbf 1 \preceq \mu^\star$ for some constant $\eta > 0$. Then, Algorithm \ref{al:dual_dual_problem} ensures that 
\begin{align*}
    \lambda  - \mathbf E[\bar \mu_k] \preceq - \frac{\eta}{2} \mathbf 1 && \forall k \ge \left(\frac{4B}{\eta^2 \sigma m } \right)^2.
\end{align*}
\label{th:tau_lemma}
\end{lemma}
Vector $\bar \mu_k$ will ensure queue stability for $k\ge \tau$. However, the queue updates require that the service vectors are schedules so that they can be implemented in practice. For that, we can decompose $\bar \mu_k$ as the convex combination of schedules, and select one randomly---as when $\lambda$ is known (see Sec.~\ref{sec:lambda_known}). We have arrived at the following theorem.

\begin{algorithm}[t!]
\caption{Schedule as You Learn (SYL)}\label{alg:new_algorithm}
\begin{algorithmic}

\State \textbf{Input:} $f$ that ensures $\lambda + \eta \mathbf{1} \preceq \mu^{*}$ (see Theorem \ref{th:main_theorem})

\While{termination condition is not met}
\State 1)~Observe packet arrivals $A_k$

\State 2)~Compute $\bar \mu_k = \frac{\sum_{i=1}^k \alpha_i \mu_i}{\sum_{i=1}^k \alpha_i}$ with Algorithm \ref{al:dual_dual_problem}
\State 3) Decompose $\bar \mu_k$, i.e.,  
\begin{align*}
\bar \mu_k = \sum_{j =1}^{|\mathcal S|} \theta_j s_j, && s_j \in \mathcal S,
\end{align*}
\State \hphantom{3)} where $\theta_j \ge 0, \sum_{ j =1}^{|\mathcal S|}  \theta_j =1$.
\State 4) Select a $S_k = s_j$ with probability $\theta_j$.
\EndWhile
\end{algorithmic}
\end{algorithm}

\begin{theorem}
Consider an arrival process $\{A_i\}_{i=1}^k$ with $\mathbf E[A_k] = \lambda$ for all $k\ge 1$. Select $S_k$ as indicated in Algorithm \ref{alg:new_algorithm}, and suppose that $\lambda +\eta \mathbf 1 \preceq \mu^\star$ for some constant $\eta > 0$, where $\mu^\star$ is the solution to the optimization in \eqref{eq:constrained_problem}. Then, the queuing system is strongly stable. 
\label{th:main_theorem}
\end{theorem}

Algorithm \ref{alg:new_algorithm} can be seen as an ``online'' version of a \emph{randomized} algorithm when $\lambda$ is known (Sec.~\ref{sec:lambda_known}). It generates randomized schedules in each time slot $k$ where the expected throughput will, eventually, be strictly larger than the mean arrival rate. Thus, the queuing system will be strongly stable (Proposition \ref{prop:queue_stability}). 

There are multiple ways to ensure $\lambda + \eta \mathbf 1 \preceq \mu^\star$. One of them is to let $\mu:= \hat \mu - \gamma \mathbf 1$ where $\hat \mu \in C$ and $\gamma \ge 0$ a slack variable, and use $f(\gamma) =  \gamma^2 -  \gamma$. The objective enforces $ \gamma$ to be strictly positive when there exists a $\hat \mu \in C$ that is strictly larger than $\lambda$, which is the case by Assumption \ref{as:slater}. Thus, we will have $\mu^\star = \hat{\mu}^\star - \gamma^\star \mathbf 1 $ where $\gamma^\star > 0$ and therefore $\lambda + \gamma^\star \mathbf 1 \preceq \hat \mu^\star$. The randomization of the schedules should be over the running average $\frac{1}{\sum_{i=1}^k \alpha_i}  \sum_{i=1}^k \alpha_i \hat \mu_i $.


To conclude, we note some limitations of the approach compared to max-weight approaches:

\begin{itemize}
    \item [(i)] Finding a decomposition of $\bar \mu_k$ is computationally more expensive than finding a maximum-weighted matching in a graph. For instance, in cross-bar switches, decomposing $\bar \mu_k$ can require computing up to $(n-1)^2 + 1$ perfect matchings/linear programs \cite{bojja2016costly, valls2021birkhoff}.  This has an impact on the frequency in which we can select schedules. 
    
    \item [(ii)] To obtain $\bar \mu_k$, we need to solve a convex optimization problem at each time slot (step 2 in Algorithm \ref{alg:new_algorithm} and implicit steps in Algorithm \ref{alg:dual_averaging_for_dual_problem}). While this can be carried out efficiently with off-the-shelf solvers, this is a step that max-weight approaches do not need to carry out. 
    
    \item [(iii)] The approach presented assumes that the network connectivity is static, whereas in max-weight approaches, the network connectivity can vary over time. Or, equivalently, when the network connectivity changes, only a fraction of the schedules in $\mathcal S$ are available at a given time. In our approach, the main technical difficulty of having a time varying connectivity is that we may not be able to express  $\bar \mu_k$ as the convex combination of the available schedules. 

\end{itemize}






\begin{figure}
\centering
\includegraphics[width=0.9\columnwidth]{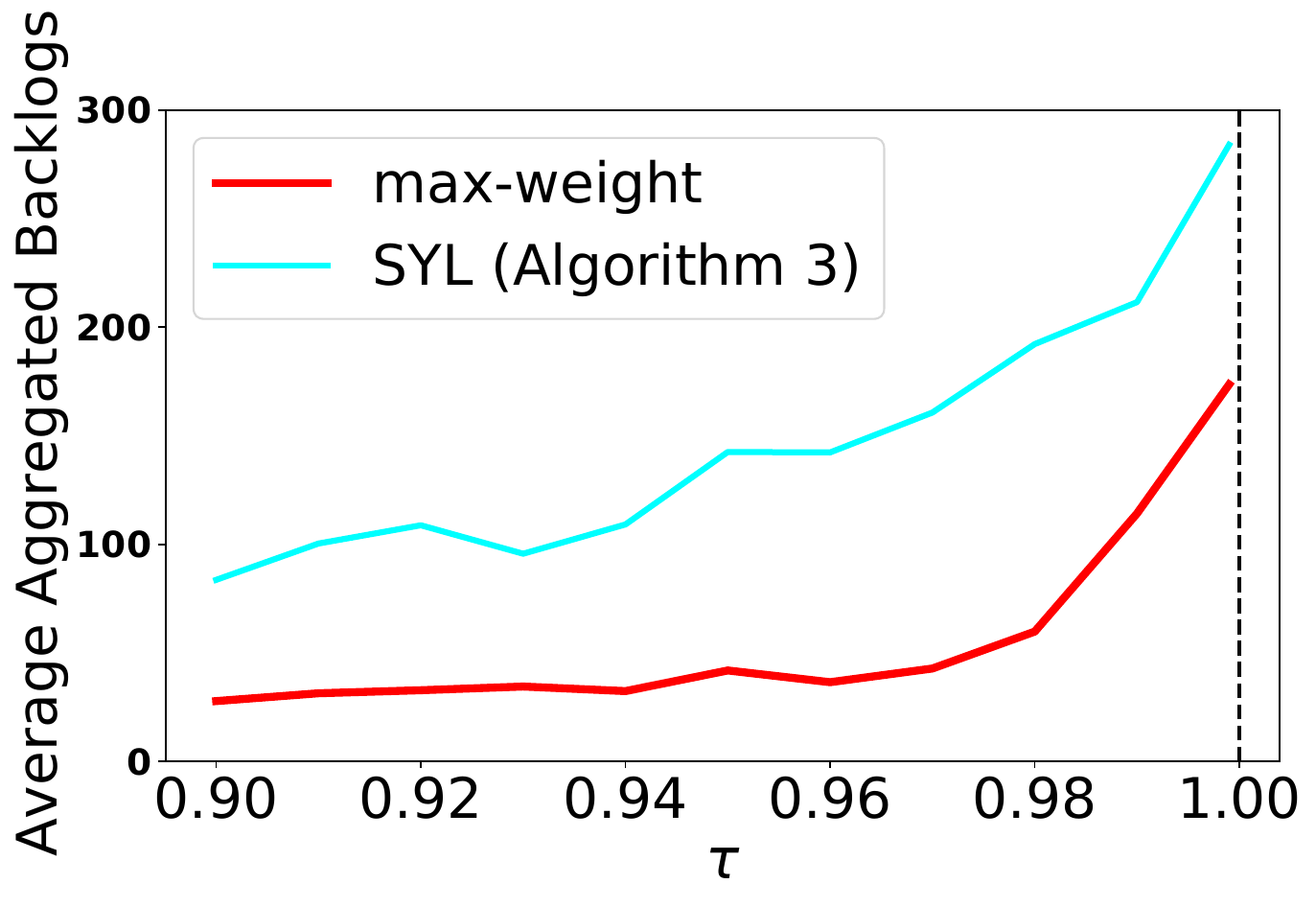}

\vspace{-0.3cm}
\caption{Aggregated backlogs in the $3 \times 3$ cross-bar switch example in Sec.~\ref{sec:num_validation_capacity} for different arrival rates $\tau(\lambda/0.9)$, where $\lambda$ is given in Eq.~\eqref{eq:lambda}.}

\label{fig:capacity_region_boundary}
\vspace{-12bp}
\end{figure}

\section{Numerical Experiments}
\label{sec:numerical_experiments}

This section presents two numerical experiments for a $3 \times 3$ cross-bar switch. Experiment A compares the performance of SYL (Algorithm \ref{alg:new_algorithm}) in terms of queue sizes. Experiment B shows an example of how we can improve the latency of a flow with a heuristic strategy based on SYL that prioritizes a flow while aiming to maintain the expected rate $\bar \mu_k$.

\subsection{Experiment A}
\label{sec:num_validation_capacity}

Consider a $3 \times 3$ cross-bar with arrivals matrix $\tau(\lambda/0.9)$, where $\lambda$ is as in Eq.~\eqref{eq:lambda}, and  $\tau \in [0,1)$ a parameter that controls the intensity of the arrivals while being within the network capacity region. The arrivals are assumed to be Bernoulli distributed with probabilities given by the matrix $\tau(\lambda/0.9)$. We run max-weight and SYL for 100k iterations and different values of $\tau$ in the range $[0.9,1)$. The simulation results are shown in Fig.~\ref{fig:capacity_region_boundary}. Observe from the figure that in both policies, the average queue backlog increases as $\tau \to 1$ and it explodes as we surpass the capacity boundary. This is the typical behaviour in max-weight approaches when the arrival rate is close the boundary of the capacity region (see, for example, \cite[Fig.~4.2]{GNT06}). However, note that max-weight has smaller backlogs sizes compared to SYL. This is because, unlike SYL, max-weight aim to minimize the \emph{total} number of packets in the queues. Minimizing the total number of packets in the queues is, however, not the best strategy if flows have different requirements in terms of latency, as we show in the following example. 




\begin{figure*}
\centering
\includegraphics[width=\textwidth]{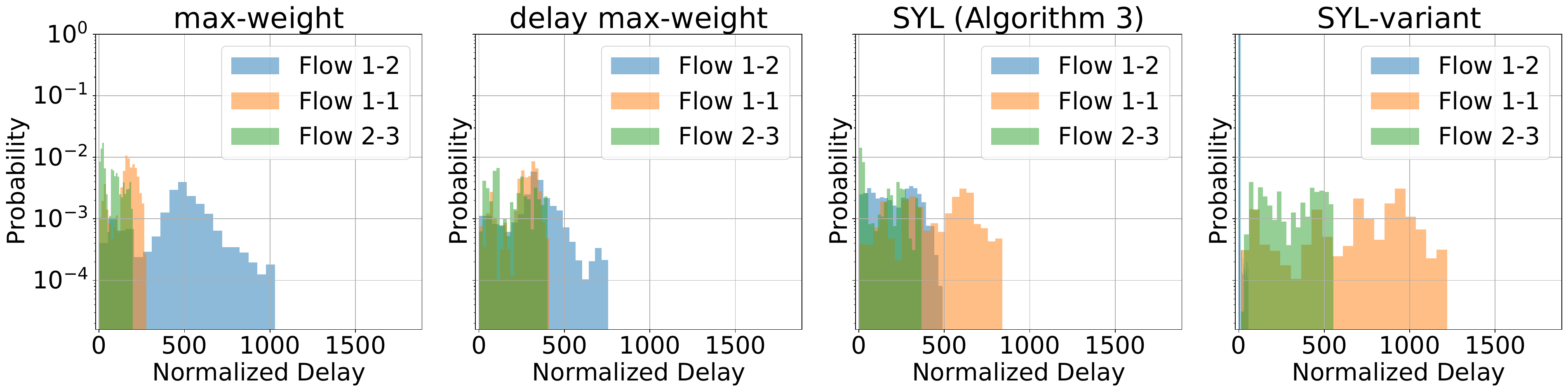}
\vspace{-0.7cm}
\caption{Comparative histograms of delay distributions across the four scheduling policies (max-weight, delay max-weight, SYL, and SYL-variant) and three different flows, showcasing the probability density of delays for pairs of nodes under three different traffic flows. For the SYL-variant we used $100$ tokens for the schedules that serve the sensitive flow 1-2 (see Sec.~\ref{sec:num_improved_performance}).}
\label{fig:delay_hist_3by3}
\vspace{-12bp}
\end{figure*}

\subsection{Experiment B}
\label{sec:num_improved_performance}


In this experiment, we consider the setting in Sec.~\ref{sec:num_validation_capacity} for a fixed $\tau = 0.98$, and suppose that the flow 1-2, going from input port to 1 to output 2, has low latency requirements.\footnote{i.e., flow 1-2 is the entry with value $0.3$ in matrix $\lambda$ in Eq.~\eqref{eq:lambda}} 
We compare four policies: max-weight \cite{TE92}, delay max-weight \cite{mekkittikul1996starvation}, SYL (Algorithm \ref{alg:new_algorithm}), and a heuristic variant of SYL. In particular, the SYL variant reserves a fixed number of ``tokens'' for schedules serving flow 1-2. In each time slot, it selects a schedule using a randomized policy as in SYL, but when sensitive flow packets arrive and tokens are available, it selects a schedule serving these packets and allocates a token to the unused schedule. If tokens are not available, it selects to the randomized schedule. In the absence of packets from the sensitive flow 1-2, if the random selection would serve it, the algorithm instead chooses the schedule with the most tokens and subtract one. 




We run the four policies for 100k iterations and show the delay distribution in Figure \ref{fig:delay_hist_3by3}. We only show in the figure flows 1-2, 1-1, and 2-3 due to space constraints, but the behavior of these flows is representative. First, observe that max-weight incurs high delays to flow 1-2 since max-weight prioritizes serving queues with larger backlogs. In other words, max-weight does not prioritize queues that grow ``slower'', which is the case of flow 1-2 since this is the flow with lowest intensity. This ``unfair'' behavior is fixed by the delay max-weight policy, which provides similar delay distributions to all flows. However, making all delay distribution similar does not guarantee low latency to the flow of interest. Finally, observe that the SYL variant provides low latency to flow 1-2 at the price of increasing the delay of the other flows. Importantly, the waiting times of the packets in the queues are bounded, which means that all the packets that get into the queues will eventually get out.




\section{Conclusions}
\label{sec:conclusions}

This paper has presented a novel approach to designing throughput optimal policies for queueing systems. The approach consists of learning an average rate that ensures queue stability, and then selecting schedules in a ``randomized'' manner that generate such a rate in expectation. An appealing characteristic of our approach is that the learning process can be offloaded to a subroutine, which allows us to focus on designing a scheduling policy for a particular system. A natural follow-up of this work is to design scheduling policies that are ``non-randomized'', and that their expected throughput ensures queue stability.

\section*{Acknowledgement}

The research work was supported by the Army Research Office MURI under the project number W911NF2110325 and by the National Science Foundation under the project number 2402862.


\bibliographystyle{IEEEtran}
\bibliography{references}

\appendix
\subsection{Proof of Proposition \ref{prop:queue_stability}}
The proof is inspired by the method of proof in max-weight policies \cite{GNT06}, but we replace max-weight by a randomized policy. To start, observe 
\begin{align*}
\| Q_{k+1} \|^2 
& = \| [Q_k + Z_k ]^+ \|^2 \\
& \le \| Q_k + Z_k \|^2 \\
& = \| Q_k \|^2 + \| Z_k \|^2 + 2 \langle Q_k, Z_k \rangle \\
& \le \| Q_k \|^2 + \sigma^2 + 2 \langle Q_k, Z_k \rangle
\end{align*}
Rearrange terms and sum from $i=1,\dots, k$
\begin{align*}
\| Q_{k+1} \|^2 \le \| Q_1 \|^2 + k \sigma^2 + 2 \sum_{i=1}^k \langle Q_i,  Z_i \rangle.
\end{align*}
Take expectations w.r.t.\ $Z_k$, and since $Q_k$ and $Z_k$ are independent,
\begin{align*}
\mathbf E[\| Q_{k+1} \|^2] \le \| Q_1 \|^2 + k \sigma^2 - 2 \eta \sum_{i=1}^k \mathbf E[ \| Q_i \|_1 ].
\end{align*}
Rearrange terms and divide by $2 \eta k$ to obtain:
\begin{align*}
\frac{1}{k} \sum_{i=1}^k \mathbf E[ \| Q_i \|_1]  \le \frac{\| Q_1 \|^2}{2k\eta} + \frac{\sigma^2}{2 \eta} .
\end{align*}
Finally, since $\| Q_1\|_2 \le \| Q_1\|_1$, take $k \to \infty$ to obtain the result. 


\subsection{Proof of Lemma \ref{th:dual_averaging}}
This proof follows the strategy explained in \cite[Sec.~2]{Nes09}. To start, let $
l_k (w) : = \sum_{i=1}^k \langle \alpha_i \nabla f(x_i), x_i - w \rangle$.
Next, observe that we can write 
\begin{align*}
l_k (w) & = \sum_{i=1}^k  \langle  \alpha_i \widetilde \nabla f(x_i),  x_i - w \rangle  + N_k \notag \\
& = \sum_{i=1}^k  \langle  \alpha_i \widetilde \nabla f(x_i), x_i \rangle + \left\langle s_{k}  , w  \right\rangle + N_k
\end{align*}
where $N_k:=  \sum_{i=1}^k  \langle  \alpha_i \xi_i,x_i - w \rangle$,  $\xi_k = \nabla f(x_k) - \widetilde \nabla f(x_k)$,  and $s_k = s_0 - \sum_{i=1}^k \alpha_i \widetilde \nabla f(x_i)$. %


We proceed to upper bound $\left\langle s_k  , w  \right\rangle$. 
Since $\phi$ is convex, by Fenchel's inequality \cite[Sec.~3.3.2]{BV04}, $\langle s_k, w \rangle \le \phi(w) + \phi^*(s_{k})$.
Thus, 
\begin{align}
  l_k(w)  & \le   \phi(w)   
 + \phi^*(s_k) + \sum_{i=1}^k  \langle \alpha_i \widetilde \nabla f(x_i), x_i \rangle + N_k
 \label{eq:ie_1}
\end{align}
Now, we upper bound the second and third terms in the right-hand-side of the last equation. 
First, observe that by the $\frac{1}{\sigma}$-smoothness of $\phi^*$ \cite{KST09}, 
\begin{align*}
& \phi^*(s_k) \\
& \le \phi^*(s_{k-1}) + \langle \nabla \phi^*(s_{k-1}), s_k - s_{k-1} \rangle +  \frac{1}{2 \sigma} \| s_k - s_{k-1} \|^2 \\
& = \phi^*(s_{k-1}) - \langle \nabla \phi^*(s_{k-1}),  \alpha_k \widetilde \nabla f(x_k) \rangle +  \frac{\alpha_k^2}{2 \sigma } \| \widetilde \nabla f(x_i) \|^2 
\end{align*}
where $s_k - s_{k-1} = - \alpha_k \widetilde \nabla f(x_k)$. 
Apply the argument recursively from $i=1,\dots,k$ to obtain
\begin{align}
 \phi^*(s_k) 
& \le \phi^*(s_{0})  - \sum_{i=1}^k \langle \nabla \phi^*(s_{i-1}),   \alpha_i \widetilde \nabla f(x_i) \rangle \notag \\
& \quad +  \frac{1}{2 \sigma} \sum_{i=1}^k  \alpha_i^2 \| \widetilde \nabla f(x_i) \|^2 
\label{eq:ie_2}
\end{align}
where $\phi^*(s_0) \le 0$ 
since $\phi$ is non-negative and $s_0 =0$. 
Also, since $x_{k} = \nabla \phi^*(s_{k-1}) = \arg \min_{u \in C} \{ \langle s_{k-1}, u \rangle - \phi(u)\}$ for all $k=1,2,\dots$ \cite[Proposition 11.3]{RW09}, we can use Eq.~\eqref{eq:ie_2} in Eq.~\eqref{eq:ie_1} to obtain
\begin{align*}
l_k(w) & \le  \phi(w)  + \frac{1}{2\sigma} \sum_{i=1}^k \alpha_i^2 \|  \widetilde \nabla f(x_i) \|^2 + N_k .
\end{align*}

Finally, note that $\mathbf E[N_k] = 0$ because the noise vector $\xi_k$ is independent of $x_k$ by assumption, i.e., $\mathbf E[\widetilde \nabla f(x_k) - \nabla f(x_k)] = \mathbf E[\xi_k] = 0$. Hence, 
\begin{align*}
\mathbf E[l_k(w)] & \le  \phi(w)  + \frac{1}{2\sigma} \sum_{i=1}^k \alpha_i^2 \|  \widetilde \nabla f(x_i) \|^2.
\end{align*}
To conclude, note that the left-hand-side of Eq.~\eqref{eq:dual_averaging_lemma} follows by the convexity of $f$.




\subsection{Proof of Lemma \ref{th:dual_dual_averaging}}

Slater's condition ensures that strong duality holds \cite[Chapter 5]{BV04}, i.e., $h^\star := \sup_{y \succeq 0} h(y) = \min_{u \in C, \lambda \preceq u} f(u) =: f^\star$. Next, since $h(y) \le h(y^\star)$ for every $y \succeq 0$ and $h(y^\star) = f^\star$, we have
\begin{align}
\sum_{i=1}^k (h(y_i) - f^\star) \le 0
\end{align}
Now, note that $h(y_k) = f(\mu_k) + \langle y_k, \lambda - \mu_k \rangle$ where $\mu_k \in \arg \min_{u \in C} \{f(u) + \langle y_k, \lambda - u \rangle \}$. Thus, rearranging terms and multiplying each term in the sum by $\alpha_k$, we have
\begin{align}
\sum_{i=1}^k \alpha_i ( f(\mu_i)  - f^\star )   \le - \sum_{i=1}^k \alpha_i \langle y_i, \lambda - \mu_i \rangle 
\end{align}
 Next, note that $-h$ is convex, and that the (sub)gradient of $-h(y_k)$ w.r.t.\ $y_k$ is  $- (\lambda - \mu_k)$. Thus, by Lemma \ref{th:dual_averaging} with $w = 0$ ($0$ in the domain of $-h$), we have 
\begin{align*}
\mathbf E \left[ \sum_{i=1}^k \alpha_i( f(\mu_i)  - f^\star)   \right]  
& \le \phi(0)   + \frac{1}{2\sigma} \sum_{i=1}^k \alpha_i^2 \| \lambda - \mu_i  \|^2 
\end{align*}
where $\phi(0) = 0$. We proceed the lower bound the left-hand-side of the last equation. By the convexity of $f$, 
\begin{align*}
& \mathbf E \left[  \sum_{i=1}^k \alpha_i (f(\mu_i)  - f^\star) \right] \ge \mathbf E \left[  \left( \sum_{i=1}^k  \alpha_i\right) (f \left( \bar \mu_k \right)  - f^\star) \right].
\end{align*}
Since $f \left( \bar \mu_k \right)  - f^\star \ge \langle \nabla f(\mu^\star) , \bar \mu_k - \mu^\star \rangle + \frac{m}{2} \left\|\bar \mu_k - \mu^\star \right\|^2 $ (by the $m$-strong convexity of $f$) and $\langle \nabla f(\mu^\star), \bar \mu_k - \mu^\star  \rangle \ge 0$ because $-\nabla f(\mu^\star) \in \mathcal N_C (x^\star) := \{ z \in \mathbf R^n \mid \langle z, \mu - \mu^\star  \rangle \le 0  \ \forall \mu \in C\}$ (see \cite[Ch.~6b]{RW09} and \cite{Lec14}), we have
\begin{align*}
\mathbf E \left[  \sum_{i=1}^k \alpha_i (f(\mu_i)  - f^\star) \right] 
\ge \frac{m}{2}  \left(\sum_{i=1}^k \alpha_i \right)     \mathbf E \left[ \left\| \bar \mu_k- \mu^\star \right\|^2\right].
\end{align*}
Finally, divide across by $\frac{m}{2} \sum_{i=1}^k \alpha_i $ to obtain a bound on $ \mathbf E [ \| \bar \mu_k- \mu^\star \|^2]$, where $\sum_{i=1}^k \alpha_i \ge \sqrt k$ and  $\sum_{i=1}^k \alpha^2_i \le \log(k) + 1 $ since $\alpha_k = \frac{1}{\sqrt{k}}$.

\subsection{Proof of Lemma \ref{th:tau_lemma}}

We will show that $\bar \mu_k$ generated by Algorithm \ref{alg:dual_averaging_for_dual_problem} satisfies $\lambda + \frac{\eta}{2} \mathbf 1 \preceq \mathbf E[\bar \mu_k]$ for all $k \ge \tau$ for some $\tau \in \{1,2,\dots\}$. Since $\sqrt x$ is concave for $x \ge 0$, we have that 
\begin{align*}
\mathbf E[\| \bar \mu_k - \ \mu^\star \| ]
= \mathbf E \left[\sqrt{\| \bar \mu_k - \ \mu^\star \|^2} \right]
\le \sqrt{ \mathbf E[\| \bar \mu_k - \ \mu^\star \|^2}.
\end{align*}
Hence, from Lemma \ref{th:dual_dual_averaging}, $
\mathbf E[\| \bar \mu_k - \ \mu^\star \| ] \le  \sqrt{\frac{B}{\sigma m} \frac{\log(k) + 1}{\sqrt{k}}}$, and therefore $\mathbf E[\| \bar \mu_k - \ \mu^\star \| ] \le \frac{\eta}{2}$ for $k \ge \left(\frac{4B}{\eta^2 \sigma m } \right)^2$.

Finally, since $\| \cdot \|_\infty \le \| \cdot\|_2$, we have that $ \mu^\star - \mathbf E[\bar \mu_k] \preceq \frac{\eta}{2} \mathbf 1$. And since $\lambda + \eta \mathbf 1 \preceq \mu^\star$ (by assumption in the lemma), it follows that
\begin{align*}
    \lambda + \eta \mathbf 1 - \mathbf E[\bar \mu_k] \preceq \frac{\eta}{2} \mathbf 1,
\end{align*}
and, therefore, $\lambda  - \mathbf E[\bar \mu_k] \preceq - \frac{\eta}{2} \mathbf 1$ for all $k \ge \left(\frac{4B}{\eta^2 \sigma m } \right)^2$.

\end{document}